\documentclass[preprint,proceedings]{rmaa}

\usepackage{psfrag,color}

\def\af{a$_{\rm f}$}
\def\Ek{E$_{\rm k}$}
\def\Es{e$_{\rm k}$}
\def\Deltat{$\Delta t_{\rm OB}$}
\def\sfr{$\dot{\Sigma}_{\star}$}
\def\lf{l$_{\rm f}$}
\newcommand\mathnew{\mathsurround=0pt}
\def\td{t$_{\rm d}$}
\def\Sigmag{$\Sigma_{\rm g}$}
\def\simov#1#2{\lower .5pt\vbox{\baselineskip0pt
    \lineskip-.5pt\ialign{$\mathnew#1\hfil##\hfil$\crcr#2\crcr\sim\crcr}}}  

\def\uf{u$_{\rm f}$}
\def\urms{u$_{\rm rms}$}

\def\grtsim{{_ >\atop{^\sim}}}

\def\lesssim{{_ <\atop{^\sim}}}
\def\msun{M$_{\odot}$}

\SetYear{2003}
\SetConfTitle{Galaxy Evolution: Theory and Observations}

\RescaleTitleLengths{0.6} 
\title{Star formation and turbulent dissipation in models
of disk galaxy evolution} 

\author{V. Avila-Reese\altaffilmark{1}, C. Firmani\altaffilmark{1,2}, and E.
        V\'azquez-Semadeni\altaffilmark{1} }

\altaffiltext{1}{Instituto de Astronom\'{\i}a, U.N.A.M., M\'exico}
\altaffiltext{2}{Osservatorio Astronomico di Brera, Italy}

\shortauthor{Avila-Reese, Firmani \& V\'azquez-Semadeni}
\shorttitle{Star formation in disk galaxies}

\fulladdresses{
\item V. Avila-Reese: Instituto de Astronom\'{\i}a, U.N.A.M., A.P. 70-264, 
04510, M\'exico, D.F., M\'exico
\item C. Firmani: Osservatorio Astronomico di Brera, via E. Bianchi 46, 
I-23807 Merate, Italy

\item E. V\'azquez-Semadeni:  Instituto de Astronom\'{\i}a, U.N.A.M., 
A.P. 3-72(Xangari), Morelia, Mich. 58089, M\'exico
}

\listofauthors{V. Avila-Reese, C. Firmani, \& E. V\'azquez-Semadeni }
\indexauthor{Avila-Reese, V.}
\indexauthor{Firmani, C.}
\indexauthor{V\'azquez-Semadeni, E.}
\abstract{The kinetic energy dissipation rate in the turbulent ISM of disk
galaxies is a key ingredient in galaxy evolution models since it determines
the effectiveness of large-scale star formation (SF) feedback. Using 
MHD simulations, we find that the ISM dissipates efficiently the turbulent
kinetic energy injected by sources of stellar nature. Thus, the SF process may
be self-regulated by an energy balance only at the level of the disk ISM.
The use of the self-regulation SF mechanism in galaxy evolutionary models, where
disks form inside growing CDM halos, allows to predict the 
SF history of disk galaxies, including the Milky Way and the solar neighborhood,
as well as the contribution of the whole population of disk galaxies to the 
cosmic SF history. The results are encouraging.}

\resumen{La tasa de disipaci\'on de energ\'{\i}a cin\'etica en el MIE turbulento
de las galaxias de disco es un ingrdiente clave para modelos de evoluci\'on
gal\'actica, ya que determina la eficiencia de la retroalimentaci\'on de la 
formaci\'on estelar (FE) de gran escala. Con ayuda de simulaciones MHD
encontramos que el MIE disipa eficientemente la energ\'{\i}a turbulenta 
inyectada por fuentes de naturaleza estelar. As\'{\i}, el proceso de FE puede ser
auto-regulado s\'olo a nivel del MIE del disco. La aplicaci\'on del mecanismo
de FE auto-regulada en modelos de evoluci\'on gal\'actica, donde los discos
se forman dentro de halos de CDM en crecimiento, permite predecir la historia
de FE de las galaxias de disco, incluyendo a la V\'{\i}a L\'actea y la 
vecindad solar, as\'{\i} como la contribuci\'on de toda la poblaci\'on de
discos a la historia de FE c\'osmica. Los resultados son alentadores. }

\addkeyword{galaxies: evolution}
\addkeyword{stars: formation}

\FinalFigure{\centering
  \includegraphics{party.ps} }

\begin{document}
\maketitle

\section{Introduction}
\label{sec:intro}

 Star formation (SF) and feedback are key processes in the evolution 
of galaxies, but probably they are the worst understood of all. The 
large-scale SF cycle in normal galaxies is  believed to be 
self-regulated by a balance between the energy injection due to SF 
(mainly SNe) and dissipation. Two main approaches have been used to 
describe the SF self-regulation in models of galaxy formation and evolution: 
{\bf (a)} the halo cooling flow-feedback approach (White \& Frenk 1991), 
{\bf (b)} the disk turbulent ISM approach (Firmani \& Tutukov
1992; 1994; Dopita \& Ryder 1994; Wang \& Silk 1994).

According to the former, the cool gas 
is reheated by the ``galaxy'' SF feedback and driven back to the 
{\it intrahalo medium} until it again cools radiatively and collapses
into the disk. This approach has been used in the semi-analytical models 
of galaxy formation
(e.g., Kauffmann, White \& Guiderdoni 1993; Cole et al. 1994). The
reheating rate depends on the halo circular velocity $V_c$: 
$\dot{M}_{rh}\propto \dot{M}_s/V_c^{\alpha}$, where $\dot{M}_*$ is the 
SF rate (SFR) and $\alpha\ge 2$. Thus, the galaxy SFR, gas fraction 
and luminosity depend on $V_c$. In these models, the disk ISM is 
virtually ignored and the SN energy injection is assumed to be as 
efficient as to reheat the cold gas up to the virial temperature
of the halo. A drawback of the model is that it predicts hot X-ray halos 
around disk galaxies much more luminous than observed (Benson et al.
2000).    

According to approach (b), SF is triggered by gravitational 
instabilities (Toomre criterion) and self-regulated by a balance between
energy injection and dissipation in the turbulent ISM in the direction
perpendicular to the disk plane:
$\gamma_{SN}\epsilon_{SN}\dot{\Sigma}_* + \dot{\Sigma}_{E,accr} =  (\Sigma_g v_g^2)
/(2t_{\rm d})$, where $\gamma_{SN}$ and $\epsilon_{SN}$ are the kinetic
energy injection efficiency of the SN into the gas and the SN energy generated 
per gram of gas transformed into stars, respectively, $\dot{\Sigma}_*$
is the surface SFR, $\dot{\Sigma}_{E,accr}$ is the surface
mass accretion rate, and $\Sigma_g$ and $v_g$ are the gas surface density
and velocity dispersion, respectively. The key parameter in 
this equation is the dissipation time \td. Now, the ISM is a 
turbulent, non-isothermal, multi-temperature flow. 
It is known that in the denser regions of the ISM, where stars form,
the cooling time is short with respect to the dynamical times (Spitzer
\& Savedoff 1950; Elmegreen 1993; V\'azquez-Semadeni et al.\ 1995,
1996). The main effect of this short cooling time is to cause the medium
to behave essentially as a polytrope, with the pressure $P$ scaling 
with density $\rho$ as $P \propto \rho^\gamma$, where $\gamma$ is an
effective polytropic exponent that determines the compressibility of
the medium
(V\'azquez-Semadeni et al.\ 1996; Scalo et al.\ 1998; Spaans \& Silk 
2000; Jenkins \& Tripp 2001). Given the medium's compressibility, the
star formation rate 
is then determined by the mean turbulent kinetic energy (which produces the
compressions), which in turn is limited by the kinetic energy
dissipation rate. Therefore, the key parameter is the {\it turbulent}
dissipation time, and a crucial question is: {\it how
efficiently does the disk ISM dissipate the \Ek\ injected by stellar
sources?}.


\section{Turbulent dissipation in ISM simulations}

To explore the ISM dissipative properties, Avila-Reese \& V\'azquez-Semadeni 
(2001) used numerical MHD simulations of turbulent compressible fluids 
resembling the ISM of disk galaxies at the 1 kpc scale (V\'azquez-Semadeni 
et al. 1995; Passot et al.\ 1995). The MHD and the internal energy conservation
equations were solved in two dimensions\footnote{We argue that
dissipation in 2-D is representative 
of that in 3-D as long as it is dominated by shocks rather than by a turbulent
cascade.} with model terms for radiative cooling
and background heating. A pseudospectral method with hyperviscocity
and periodic boundary conditions was used. The way in which \Ek\ is injected
to the fluid is crucial: it should resemble ``stellar'' sources like
SN remnants and expanding HII regions. That is, the energy sources are
discrete, generally small in comparison with the characteristic
scales of the global ISM, and with filling factors typically $\ll
1$. The forcing in the simulations 
was generated by ``winds'': the gas around the SF regions received
an acceleration \af, directed radially away, thus producing an evolving
velocity profile with characteristic forcing velocity \uf\ and radius
\lf. The characteristic energy injected is: \Es $\propto$ \uf \af
\lf$^2$ \Deltat,  
where, \Deltat\ is the source lifetime (assumed
constant and =6.8 Myr). The two independent input parameters that define
the local energy injection process are \lf\ and \Es\ (or \uf); \af\ is
already a response of the flow. There  
is a third initial parameter related to the global energy injection:
the surface SFR, \sfr. 

We performed a sizeable number of simulations to explore how the 
dissipative properties of the turbulent ISM depend on the forcing 
parameters: \lf, \Es, and \sfr. In these simulations, the
sources are placed randomly in space with a fixed probability, which 
determines the \sfr. 

The simulations show that the kinetic energy injection and 
dissipation rates are always very close to each other, suggesting 
that most of the dissipation occurs at or near the sources, where
shocks are common, and that, for practical purposes, the injection and
dissipation times are equal. In general, a flow with physical parameters
close to the Galactic disk at the solar neighborhood dissipates its 
turbulent energy rapidly, in approximately 15--20 Myr. A more realistic
simulation, where the SF is density directed, gives \td$\approx 18$ Myr
(in this simulation, the input sources tend to cluster due to supershell
formation). We also showed that, in terms of measurable properties of
the ISM, \td\  can be expressed
as: \td $\gtrsim \langle$\Sigmag$\rangle$ \urms$^2/$(\Es \ \sfr),
where $\langle$\Sigmag$\rangle$ is the average gas surface 
density, and the inequality arises because some overlapping of the forcing
regions may occur.  Since \sfr$\propto$ \Sigmag$^n$ (Schmidt law),
\td $\propto$ \Sigmag$^{(n-1)}(r)$. Thus, \td\ is expected to decrease
with Galactocentric radius. 

Given the discreteness and low filling
factor of the energy sources, the turbulence ``propagates'' away from
them into the general ISM, reaching out to farther distances for lower \td.
The velocity dispersion, \urms, of the residual 
turbulence is found to decay roughly inversely 
with distance to the source. It was also found that the dominant
parameter determining \td\ is \lf, while the energy per source and
the SFR are not very relevant (\td $\propto
$\lf$^{0.84}/$(\Es \ \sfr)$^{0.12}$). Thus, the production of 
very large active turbulent zones (able to inject energy out from the disk 
into the halo, for example), which occurs at large \td, requires large
values of the source characteristic size \lf, i.e., that the forcing regions 
are themselves large (supershells). These supershells could arise as a 
consequence of strong clustering and self-propagating SF. 
This might be the rule for starburst galaxies (non-stationary SF), but 
generally the exception in normal disk galaxies.

Our results, if applicable in the vertical direction, imply that 
the turbulent \Ek\ produced in the disk plane is mostly dissipated 
inside the disk ISM. The turbulent motions will propagate up to distances 
close to the observed HI disk semi-thickness, 
and will not be able to reheat and drive back the gas into the halo.
The shock-dissipated turbulent energy transforms probably into thermal
energy, which could be a source of high-energy photons able to ionize the 
low-density extraplanar medium (up to a few kpcs from the plane).

\RescaleFloatLengths{0.7} 
\begin{figure}[!t]
\vspace{7.6cm} \includegraphics{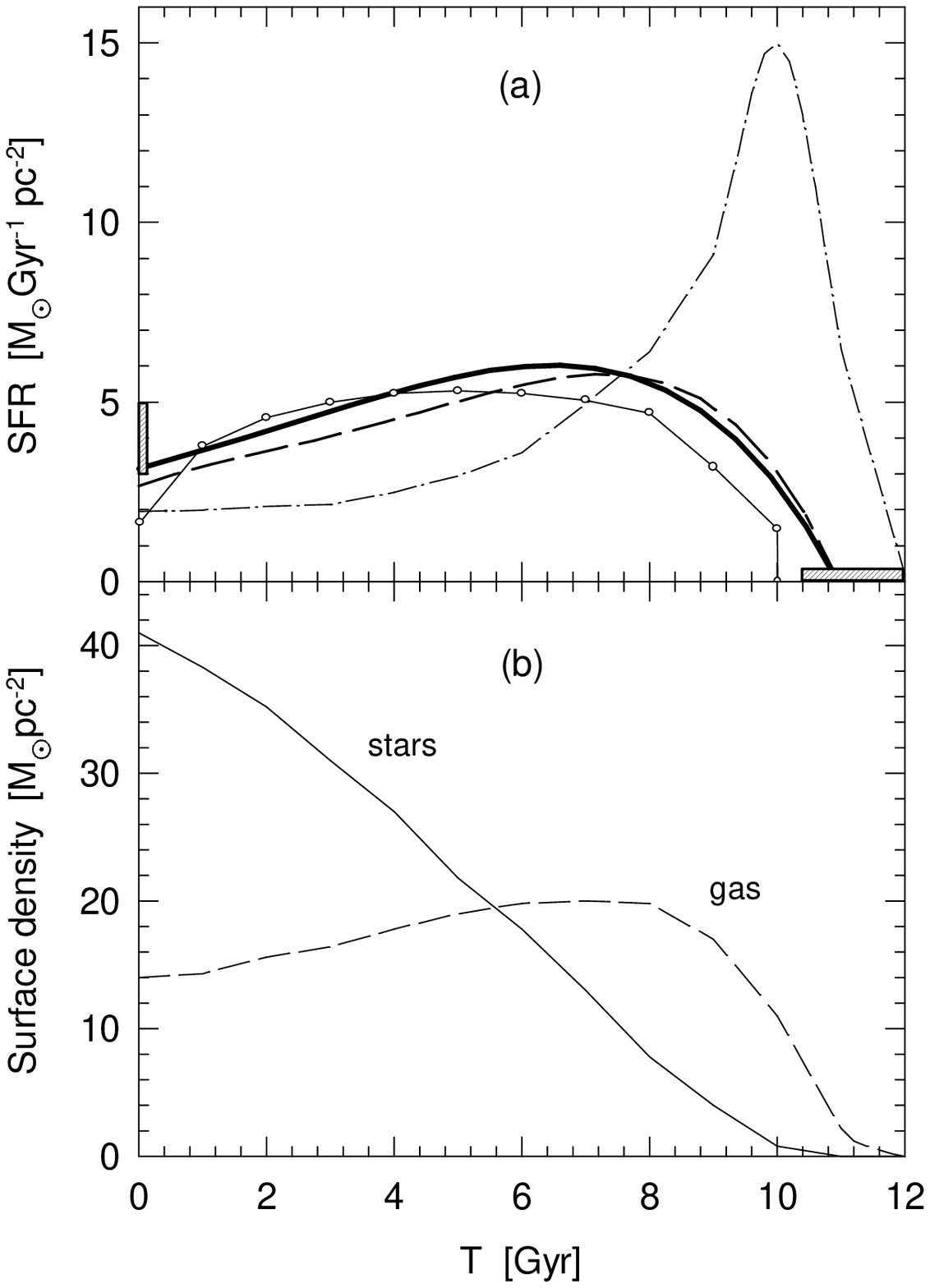}
  \caption{{\it a).} SF history in the solar neighborhood:
predicted by our model (thick line) and infered from observations 
(jointed circles). Dashed line is for a MW model
with a soft core in the halo. Dot-dashed line is the predicted gas infall 
rate at $R_{\odot}$ . The vertical and horizontal shaded boxes 
are observational estimates for the present-day SFR and for the age
of the solar neighborhood, respectively. {\it b).} Predicted evolution
of the star (solid line) and gas (dashed line) surface densities at the 
solar neighborhood.}
  \label{fig:plot1}
\end{figure}

\RescaleSecLengths{0.8} 
\section{The star formation history}

A scheme of SF triggered by disk gravitational instabilities and 
self-regulated by the SN energy input and the
turbulent energy dissipation was applied by Firmani, Hern\'andez \& Gallagher
(1996) to galaxy evolution models. A Schmidt law with index $\lesssim 2$
along a major portion of the disks was predicted. For closed models, even
with the lowest disk surface densities, the SF timescales result shorter
than 5 Gyr and the integral $(B-V)$ colors redder than 0.9. Gas infall is necessary.
This infall may be provided by the cosmological mass aggregation predicted in a 
hierarchical CDM scenario, where disks form inside-out.

Self-consistent models of disk galaxy formation and evolution within
growing CDM halos, which use the disk self-regulated SF approach, indeed predict
realistic integral and radial color indices, gas fractions, surface brightness 
profiles as well as realistic dynamical properties and correlations among
the galaxy parameters  (see Firmani \& Avila-Reese, this 
volume). In these models, the turbulent dissipation time was calculated
as \td$(r) = a (2\Omega(r))/\kappa(r)^2$,
where $\Omega(r)$ and $\kappa(r)$ are the angular velocity and epicyclic
frequency at radius $r$, and $a$ is a constant close to unity (Firmani et al.
1996). The formula above was derived from a simple dynamical analysis.
The MHD simulations of the turbulent ISM presented in \S 2 offer the
posibility to fix the value of $a$. Normalizing to the Galactic disk
at the solar radius, we find that $a\approx 1.5-2$. The dependence
of \td\ on the radius in the formula above agrees with an inference
from the simulations of such a dependence (see above).

The SF histories of most of the model galaxies have a broad maximum
at $z\sim 1-2.5$, and then the SFR decays by factors of 1.5-4
until $z=0$ (Avila-Reese \& Firmani 2001). The SF history is driven
by the halo angular momentum and by the halo mass aggregation history;
in contrast with the halo cooling-flow model, there is not a 
significative dependence on mass or circular velocity. The angular
momentum determines the disk surface density. The higher the density, the
more efficient is the SFR. The mass aggregation history determines the 
gas infall rate; for mass assembly histories of the halo more extended
in time, the late gas accretion rate is higher, and therefore, the SFR
is higher at $z\approx 0$.    
Thus, HSB and/or red galaxies have a maximum SFR at earlier epochs
and a faster decline towards the present than LSB and/or blue galaxies.
The latter may even have an increasing SFR at $z=0$. When comparing
the SFR of high-redshift disk galaxy populations with the present one,
it is important to take into account these differences of the SF history
on galaxy surface brightness and color and the selection effects. 

\RescaleSecLengths{0.7}
\subsection{Milky Way (MW) and the solar neighborhood}

The global and local SF histories of a MW model can be predicted.  
For a $\Lambda$CDM cosmology, the MW model was constructed using 
a virial mass of $2.8 \ 10^{12}$\msun, a baryon fraction of 0.02, 
the average halo mass aggregation history, and a spin parameter 
$\lambda =0.02$. The surface brightness profiles and scale radii 
(bands $B$ and $K$), the global gas and stellar mass fractions, and the 
integral colors are well reproduced (Hern\'andez, Avila-Reese \& Firmani
2001). The rotation curve decomposition is also reproduced, albeit with
an excess of dark matter within the solar radius, $R_{\odot}=8.5$ kpc. At 
$R_{\odot}$, 
the gas and stellar surface densities, as well as the 
$B-K$ color at the present epoch, agree rather well with observational 
estimates (Fig. 1b). 

The present-day {\it global} SFR of the MW model is 2.9 \msun yr$^{-1}$; it
increases slightly towards the past, attaining a maximum at 9.8 Gyr ago 
($z\approx 1.3$).  The {\it local} surface SFH at $R_{\odot}$ is shown in 
Figure 1a (thick solid line, in unities of \msun yr$^{-1}$ pc$^{-2}$). It 
starts to 
increase 11 Gyr ago (roughly when the disk forms at this radius), attains a 
maximum $\sim 6-7$ Gyr ago, and then decreases by a factor of two to the 
present epoch. By using a sample of field stars from the Hipparcos catalogue
and comparing it with synthetic colour-magnitude diagrams, the SFR per
unit of volume at $R_{\odot}$ was infered by Bertelli \& Nasi (2001). 
Hern\'andez et al. (2001) converted this SFR to a surface SFR  (open 
circles in Fig. 1a). The agreement between observations and model is 
encouraging. The introduction of a soft core in the CDM halo (to 
fit better the observed rotation curve) does not influence significantly
on the predicted SFR history (dashed line in Fig. 1a). We notice that
the observational inference describes only the general trend of the SFH
over the total life of the system, and not its detailed shape, which may
be discontinuous. The model results were smoothed in time to facilitate
the comparison. Interactions with satellites can introduce bursts
of SF (e.g., Rocha-Pinto, Maciel \& Scalo 2000; Kauffmann, Charlot \& 
Balogh 2000).

\subsection{Integral disk population SF history}

Figure 2 reproduces a compilation of data for the 
integral SFR in the universe at different redshifts presented by 
Kravtsov \& Yepes (2000). In this Fig., the integral SF history due 
to only (model) disks is also shown (thick dashed line). At low 
redshifts the observed SFR decreases more rapidly
than models. Models give actually an upper limit for the disk contribution
to the cosmic SF history, because they do not take into account 
neither the truncation of the gas infall in the massive galaxies due to 
large gas cooling times and/or external factors, nor the fact that 
at late epochs some of the disk accreting material might already be in 
form of stars (satellite or companion galaxies). At $z\grtsim 2$, most of the 
data collected by Kravtsov \& Yepes, as well as more recent estimates 
(e.g., Nandra et al. 2002; shaded rectangle), show 
that the SFR remains roughly constant, while the (model) disk contribution
decreases. A recent re-analysis of the {\it HST} deep field by Lanzetta et al. 
(2002) shows that the SFR {\it even increases significantly} at redshifts 
larger than 2. 
The model disk galaxy population ($\Lambda$CDM cosmology) seems to be
able to explain the observed cosmic SF history inferences until $z\approx 2$, 
modulo the strong fall since $z\approx 0.5$. At epochs earlier than $z\approx 2$, 
the contribution of disk galaxies to the cosmic SFR slightly and continuosly
decreases.
 
\RescaleFloatLengths{0.7}
\begin{figure}
  \includegraphics[width=\columnwidth]{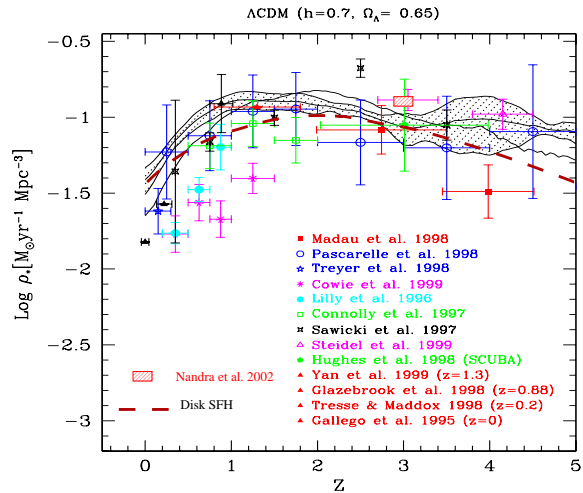}
  \caption{Dust corrected estimates of the cosmic SFR at different redshifts 
(points) from UV and H$\alpha$ observations as collected by Kravtsov \& Yepes 
(2000), and from X-ray observations of $z\sim 3$ LBGs. The data points 
have been converted to the $\Lambda$CDM cosmology. Dashed 
region is from the Kravtsov \& Yepes gasdynamical simulations, while the 
dashed solid line is our model estimate of the contribution to the cosmic 
SF history by stars formed in the disks (the $\Lambda$CDM model was used).}
  \label{fig:plot1}
\end{figure}

\RescaleSecLengths{0.7}
\section{Conclusions}

The SF process is a key ingredient for models of galaxy formation and
evolution. For disk galaxies, stationary (non-bursting) and self-regulated
SF seems to be a reasonable description. In this description, the large-scale
SF is triggered by gravitational instabilities of the gaseous disk (Toomre
criterion) and self-regulated by an energy balance in the ISM perpendicular
to the disk plane. The key factor in this energy balance is the ability
of the turbulent ISM to dissipate the kinetic energy; the turbulent 
dissipation time determines the effectiveness of large-scale SF feedback. 

Our MHD, nonisothermal simulations of the ISM including realistic, stellar-like 
forcing, show that turbulent 
energy is dissipated rapidly, mostly at or near the sources, where shocks
are common. The dissipation time \td\ is found to increase with the forcing
scale \lf\ and depends only very weakly on \Es\ and \sfr. For properties
of the ISM close to those at the solar neighborhood,
\td$\approx 15-20$ Myr. Forced and decaying turbulent regimes coexist
within the same flow. The rms velocity dispersion of the ``residual''
turbulence decays roughly inversely with distance from the injection sources.
Turbulent motions produced near the disk plane will propagate
only to distances of the order of the observed HI disk semi-thickness. 
This is consistent with the disk self-regulated SF scheme mentioned 
above, where the HI thickness results from the balance between kinetic
energy injection and its dissipation rate.

The application of this scheme to self-consistent models of disk galaxy
formation and evolution within growing $\Lambda$CDM halos leads to 
rather succesful predictions for the SFR, surface brightness, and color
radial profiles of the disk galaxy population. The main drivers of the SF history
are both the gas infall rate determined by the cosmological mass aggregation
rate, and the disk surface density determined by the halo angular momentum.
The SFR depends very weakly on mass. The SF history predicted for the 
solar neighborhood is in excellent agreement with observational inferences.
The predicted integral disk SF history describes well the observed
cosmic SF history until $z= 2-3$. At earlier epochs, the
disk population SFR decreases while the observations show that the
cosmic SFR remains constant or even increases, suggesting that at early 
epochs most of the UV and X-ray radiation is not produced by 
SF in disks.

\acknowledgments

This work was supported by CONACyT grant 33776-E to V.A.


\begin{thebibliography}

\bibitem{} Avila-Reese, V., \& Firmani, C. 2001, RevMexA\&A (CS), 10, 97
\bibitem{} Avila-Reese, V. \& V\'azquez-Semadeni, E. 2001, \apj, 553, 645
\bibitem{} Benson, A.J., Bower, R.G., Frenk, C.S., \& White, S.D.M. 
    2000, \mnras, 514, 357
\bibitem{}Bertelli G., \& Nasi E., 2001, AJ, 121, 1013.
\bibitem{}  Cole, S., Aragon-Salamanca, A., Frenk, C.S., Navarro, 
\linebreak \adjustfinalcols J., \& Zepf, S. 1994, \mnras, 271, 781
\bibitem{}  Dopita, M.A., \& Ryder, S.D. 1994, \apj, 430, 163
\bibitem{}  Firmani, C., \& Tutukov, A.V. 1992, \aap, 264, 37
\bibitem{}  \_\_\_\_\_\_\_. 1994, \aap, 288, 713
\bibitem{}  Firmani, C., Hern\'{a}ndez, X., \& Gallagher, J. 1996, \aap, 308,
403
\bibitem{} Hern\'andez, X., Avila-Reese, V. \& Firmani, C. 2001,
\mnras, 327, 329
\bibitem{} Jenkins, E.B. \& Tripp, T.M., 2001, ApJS, 137, 297
\bibitem{} Kauffmann, G., Charlot, S., \& Balogh, M.L. 2001, (astro-ph/0103130)
\bibitem{} Kauffmann, G., White, S.D.M., \& Guiderdoni, B. 1993, \mnras,
264, 201
\bibitem{} Kravtsov, A.V. \& Yepes, G. 2000, \mnras, 318, 227
\bibitem{} Lanzetta, K. M., Yahata, N., Pascarelle, S., Chen, H., 
Fern\'andez-Soto, A. 2002, \apj, 570, 492
\bibitem{} Nandra, K, et al. 2002,  \apj, 576, 625
\bibitem{} Passot, T., V\'azquez-Semadeni, E. \& Pouquet A. 1995,
\apj, 455, 536
\bibitem{} Rocha-Pinto H., Maciel W.J., Scalo J., \& Flynn, C. 2000,
A\&A, 358, 869. 
\bibitem{} Scalo, J., V\'azquez-Semadeni, E., Chappell, D.\ \& Passot,
T., 1998, \apj, 504, 835
\bibitem{} Spaans, M.\& Silk, J. 2000, \apj, 538, 115
\bibitem{} V\'azquez-Semadeni, E., Passot, T., \& Pouquet A. 1995, \apj,
441, 702
\bibitem{} \_\_\_\_\_\_\_. 1996, \apj, 473, 881
\bibitem{} Wang, B., \& Silk, J. 1994, \apj, 427, 759
\bibitem{}  White, S.D.M, \& Frenk, C.S. 1991, \apj, 379, 52

\end{thebibliography}
\end{document}